\documentclass[12pt,a4paper]{article}
\usepackage[text={16.5cm,23cm},centering]{geometry}
\usepackage{amsmath}
\usepackage{amsthm}
\usepackage[utf8]{inputenc}
\usepackage{tgtermes,tgheros,tgcursor}
\usepackage[varg]{newtxmath}
\usepackage{graphicx}
\usepackage{hyperref}
\hypersetup{%
 colorlinks=true,%
 linkcolor=blue,
 citecolor=blue,
}
\usepackage{xspace}

\numberwithin{equation}{section}

\newcommand{\imaginary}{\mathrm{i}}

\newcommand{\Zb}{\mathbb{Z}}
\DeclareMathOperator*{\Tr}{\mathrm{Tr}}
\newcommand{\SU}[2]{\mathrm{SU}(#1)_{#2}}
\newcommand{\Unitary}[2]{\mathrm{U}(#1)_{#2}}
\newcommand{\SO}[1]{\mathrm{SO}(N)_{#1}}
\newcommand{\Uone}{\mathrm{U}(1)}
\newcommand{\Ubaryon}{\mathrm{U}(1)_{B}}
\newcommand{\Nsqrt}{\mathrm{e}^{2\pi\imaginary/N}}
\newcommand{\Nsqinv}{\mathrm{e}^{-2\pi\imaginary/N}}
\newcommand{\halfNsqinv}{\mathrm{e}^{-\pi\imaginary/N}}
\newcommand{\diag}{\mathrm{diag}}
\newcommand{\exdel}{\mathrm{d}}
\newcommand{\oddfermion}{\Psi}
\newcommand{\evenboson}{\Phi_{\mathrm{even}}}
\newcommand{\oddboson}{\Phi_{\mathrm{odd}}}
\newcommand{\Tssb}{T_{\mathrm{chiral}}}
\newcommand{\Tconf}{T_{\mathrm{confinement}}}
\newcommand{\Zbr}{(\mathbb{Z}_{2})_{\mathcal{R}}}
\newcommand{\Zbc}{(\mathbb{Z}_{N})_{c_{1}}}
\newcommand{\axial}{\mathrm{axial}}
\newcommand{\CS}{\mathrm{CS}}

\begin{document}
\begin{center}
  \begin{flushright}
    OU-HET 1157
  \end{flushright}
  \vspace{8ex}
  {\Large \bfseries \boldmath Phase structure of linear quiver gauge theories from anomaly matching}\\
  \vspace{4ex}
  {\Large Okuto Morikawa, Hiroki Wada, Satoshi Yamaguchi}\\
  \vspace{2ex}
  {\itshape Department of Physics, 
  Osaka University, Toyonaka, Osaka 560-0043, Japan}\\
  \vspace{1ex}
  \texttt{o-morikawa@het.phys.sci.osaka-u.ac.jp}\\
  \texttt{hwada@het.phys.sci.osaka-u.ac.jp}\\
  \texttt{yamaguch@het.phys.sci.osaka-u.ac.jp}\\
  \begin{abstract}
    We consider the phase structure of the linear quiver gauge theory, using the 't~Hooft anomaly matching condition.
    This theory is characterized by the length $K$ of the quiver diagram.
    When $K$ is even, the symmetry and its anomaly are the same as those of massless QCD.
    Therefore, one can expect that the spontaneous symmetry breaking similar to the chiral symmetry breaking occurs.
    On the other hand, when $K$ is odd, the anomaly matching condition is satisfied by the massless composite fermions.
    We also consider the thermal partition function under the twisted boundary conditions.
    When $K$ is even, from the anomaly at finite temperature, we estimate the relation between the critical temperatures associated with the confinement/deconfinement and the breaking of the global symmetry.
    Finally we discuss the anomaly matching at finite temperature when $K$ is odd.
  \end{abstract}
\end{center}

\vspace{4ex}
\section{Introduction and summary}
The symmetry of physical systems provides quite useful tools to extract non-trivial information on the dynamics of quantum field theories.
The global symmetry, in particular, would constrain the phase structure of strongly coupled theories, notably through the use of 't~Hooft anomaly matching conditions~\cite{tHooft:1979rat}.
It is traditional to consider symmetries acting on \emph{local} operators, while recently there are various analyses of symmetries which focus on operators extended in spacetime such as the Wilson loops and the 't~Hooft loops; these are called the higher form symmetries~\cite{Gaiotto:2014kfa}.
Then, a $p$-dimensional operator may have the charge generated by the corresponding higher form symmetry, called the $p$-form symmetry.
We now note that anomalies with regard to the $p$-form symmetry should also satisfy the 't~Hooft anomaly matching condition.
In the $\SU{N}{}$ pure Yang-Mills theory, for instance, there exists the mixed 't~Hooft anomaly between the time reversal symmetry and the $1$-form $\Zb_{N}$ symmetry, i.e., the center symmetry. From this fact, one finds that the trivially gapped phase at~$\theta=\pi$ is ruled out~\cite{Gaiotto:2017yup}.
It is then natural to ask what about the 't~Hooft anomaly matching for higher form symmetries in $4$-dimensional gauge theories with matter fields such as QCD~\cite{Tanizaki:2017bam,Shimizu:2017asf,Kitano:2017jng,Tanizaki:2017qhf,Yamazaki:2017dra,Tanizaki:2017mtm,Cherman:2017dwt,Draper:2018mpj,Anber:2018iof,Cordova:2018acb,Anber:2018jdf,Yamaguchi:2018xse,Anber:2018xek,Armoni:2018bga,Bolognesi:2019fej,Sulejmanpasic:2020zfs,Bolognesi:2021yni,Bolognesi:2021hmg}.\footnote{See~\cite{McGreevy:2022oyu,Cordova:2022ruw} for reviews and references therein.}

\begin{figure}[thbp]
    \begin{minipage}[]{0.29\textwidth}
    \centering
    \includegraphics[width=2.6cm]{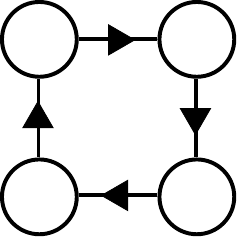}
  \end{minipage}
    \begin{minipage}[]{0.69\textwidth}
    \centering
    \includegraphics[width=6.8cm]{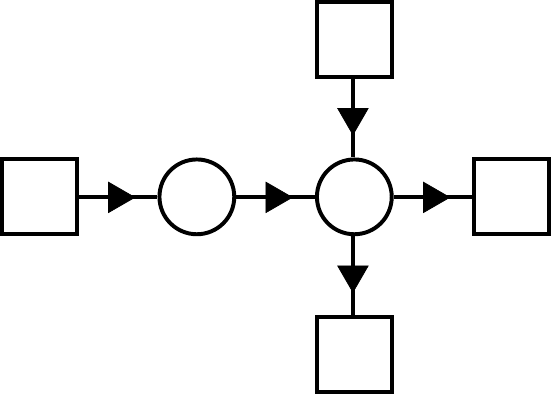}
  \end{minipage}
  \caption{Quiver diagrams}\label{fig_general_quiver}
\end{figure}

We consider the $4$-dimensional quiver gauge theories, which are depicted by the corresponding quiver diagrams as~fig.~\ref{fig_general_quiver}.
In these figures, the round nodes stand for the $\SU{N}{}$ gauge group, while the square nodes are the $\SU{N}{}$ \emph{global} symmetry group.
An edge with arrow denotes one Weyl fermion in the bifundamental representation of the two connected nodes, $\SU{N}{}\times\SU{N}{}$.
This theory is generally a chiral gauge theory.
For an original motivation, Georgi~\cite{Georgi:1985hf} has investigated such quiver gauge theories as models of composite fermions which appear in the strong coupling regime.
We are also motivated toward the so-called dimensional deconstruction~\cite{Arkani-Hamed:2001kyx}.
Since these theories are strongly coupled, however, it is difficult to examine the low-energy dynamics and their phase structures.
The authors of~\cite{Sulejmanpasic:2020zfs} have studied the \emph{circular} quiver gauge theories such as the left panel of~fig.~\ref{fig_general_quiver}, whose diagrams do not have any ends, through the analyses of the anomaly matching for the $1$-form symmetry; one expects that there are $N$ degenerate vacua in the $\SU{N}{}\times\dots\times\SU{N}{}$ gauge theory.

\begin{figure}[thbp]
    \begin{minipage}[b]{0.49\textwidth}
    \centering
    \includegraphics[width=6.3cm]{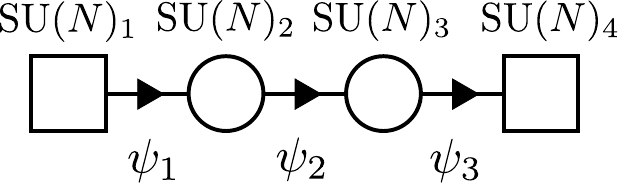}
  \end{minipage}
    \begin{minipage}[b]{0.49\textwidth}
    \centering
    \includegraphics[width=8.1cm]{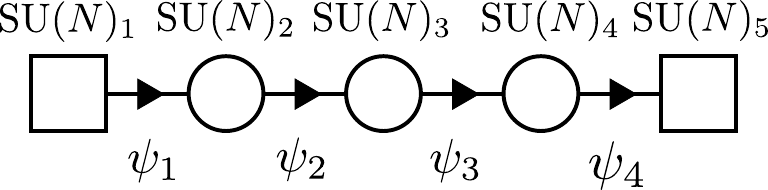}
  \end{minipage}
  \caption{Linear quiver diagrams}\label{fig_quiver}
\end{figure}

The linear quiver gauge theories represented in~fig.~\ref{fig_quiver} possess a quite different behavior from the circular quiver gauge theories at the infrared regime; they are generally gapless, while the circular ones are gapped.
We can see this behavior by considering the fact that two ends of a linear quiver diagram, i.e., the global $\SU{N}{}^2$ group, give rise to the perturbative anomaly.
Note that the idea of the dimensional deconstruction implies that the line spanned by the edges provides an extra dimension and the two square nodes at the ends correspond to boundaries.
It is thus interesting that the linear quiver gauge theories are quite similar to topological materials.
We know that gapless modes are localized at boundaries (or defects) of topological materials~\cite{Callan:1984sa}, while the system is gapped without boundaries.
This similarity may be understood from the perspective of the 't~Hooft anomaly matching in a more robust way.

In this paper, we focus on the linear quiver gauge theory explicitly depicted in~fig.~\ref{fig_quiver}.
The diagram with~$K$ edges provides the $\SU{N}{}^{K-1}$ gauge theory with matter fields of $K$~bifundamental Weyl fermions.
When~$K\ge3$, this is a chiral theory and the mass terms of fermions are prohibited.

We utilize the anomaly matching to study the linear quiver gauge theory at the infrared regime.
We take into account not only perturbative anomalies but also anomalies involving the $1$-form symmetry.
The anomalies of the theory are different depending on whether $K$ is even or odd.
Thus, the behavior of the dynamics at the infrared regime is also quite different.
When $K$ is even, the symmetry and the anomalies of the theory are the same as massless QCD.
Thus, the anomalies can be matched by the spontaneous symmetry breaking similar to the chiral symmetry breaking.
On the other hand, when $K$ is odd, it is most likely that the  massless composite fermions appear at the infrared regime \cite{Georgi:1985hf}.
In this paper, we show that the composite-fermions scenario is consistent with the anomaly matching condition involving the $1$-form symmetry.
We also point out that the anomalies for odd $K$ can be matched by the spontaneous symmetry breaking of the global symmetry.

We also discuss the phase structure of the linear quiver gauge theory at finite temperature.
The phase structure of the pure Yang-Mills theory is constrained by the anomaly involving the center symmetry \cite{Gaiotto:2017yup}.
Although in general gauge theories with matter fields such as massless QCD do not have the center symmetry, the anomaly matching still works by considering the thermal partition function under the twisted boundary conditions along the thermal circle \cite{Shimizu:2017asf,Tanizaki:2017qhf,Tanizaki:2017mtm,Yonekura:2019vyz}.
In this paper, we apply this method to the liner quiver gauge theory.
When $K$ is even, we argue the relation between the critical temperatures associated with the confinement/deconfinement and the breaking of the global symmetry.
When $K$ is odd, we check that the above two possibilities at the infrared regime are also consistent with the anomaly at finite temperature.

The construction of this paper is as follows.
In section \ref{sec_setup}, we give the detail of the linear quiver gauge theories.
We also deduce their global symmetry.
In section \ref{sec_even}, we study the phase structure of the theory with even $K$.
Section \ref{sec_odd} is devoted to the analyses for the theory with odd $K$.

\section{Linear quiver gauge theories and their global symmetries}\label{sec_setup}
\subsection{Setup}
We consider the linear quiver gauge theory with the gauge group
\begin{align}
        \SU{N}{}^{K-1}=\SU{N}{2}\times\SU{N}{3}\times\cdots\times\SU{N}{K}.
    \end{align}
It contains Weyl fermions $\psi_{i}$ in the bifundamental representation of $\SU{N}{i}\times\SU{N}{i+1}$ for $i=1$, $2$, \dots, $K$. Here, $\SU{N}{1}$ and $\SU{N}{K+1}$ are global symmetries.
The gauge field of $\SU{N}{i}$ is denoted by $a_{i}$.
The field strength $f_{i}$ is defined by $f_{i}=\exdel a_{i}+a_{i}^{2}$.
Gauge fields and field strengths are taken to be anti-hermitian throughout the present paper unless otherwise noted.
The action in the Euclidean signature is written as
    \begin{align}
        S=\sum_{i=1}^{K}\int\left(\bar{\psi}_{i}\bar{\sigma}_{\mu}D_{\mu}\psi_{i}\right)
        +\sum_{i=2}^{K}\int\left(-\frac{1}{h_{i}^{2}}\Tr f_{i}\wedge\ast f_{i}\right),
    \end{align}
where $h_{i}$ are the gauge coupling constants.
The covariant derivative on the fermions is given by
    \begin{align}
    D\psi_{1}&=\exdel\psi_{1}-\psi_{1}a_{2},\\
    D\psi_{i}&=\exdel\psi_{i}+a_{i}\psi_{i}-\psi_{i}a_{i+1}\qquad(i=2,\,\dots,\,K-1),\\
    D\psi_{K}&=\exdel\psi_{K}+a_{K}\psi_{K}.
    \end{align}
The gauge transformations are given by
    \begin{align}
    a_{i}&\mapsto a_{i}'=g_{i}a_{i}g_{i}^{-1}+g_{i}\exdel g_{i}^{-1}\qquad(i=2,\,3,\,\dots,\,K),
    \end{align}
and
    \begin{align}
    \psi_{1}&\mapsto\psi_{1}'=\psi_{1}g_{2}^{-1},\\
    \psi_{i}&\mapsto\psi_{i}'=g_{i}\psi_{i}g_{i+1}^{-1}\qquad(i=2,\,3,\,\dots,\,K-1),\\
    \psi_{K}&\mapsto\psi_{K}'=g_{K}\psi_{K},
    \end{align}
where $g_{i}$ are the $\SU{N}{i}$-valued gauge transformations.

\subsection{Global symmetries}
We show that the symmetry group which faithfully acts on the fermions is given by
    \begin{align}\label{total_sym}
        \frac{\SU{N}{1}\times\SU{N}{2}\times\cdots\times\SU{N}{K}\times\SU{N}{K+1}\times\Uone}{\Zb_{N}\times\Zb_{N}},
    \end{align}
including the gauge as well as the global symmetry groups.
The above $\Uone$ acts on the fermions as
    \begin{align}\label{uone_trans}
        \psi_{i}\mapsto\psi_{i}'=\mathrm{e}^{\imaginary\theta(-1)^{i}}\psi_{i}\qquad(i=1,\,\dots,\,K),
    \end{align}
with $\theta\sim\theta+2\pi$.
The $\SU{N}{1}\times\SU{N}{K+1}$ symmetry transforms $\psi_{1}$ and $\psi_{K}$ as
    \begin{align}
        \psi_{1}&\mapsto\psi_{1}'=g_{1}\psi_{1},\\
        \psi_{K}&\mapsto\psi_{K}'=\psi_{K}g_{K+1}^{-1},
    \end{align}
where $(g_{1},\,g_{K+1})\in\SU{N}{1}\times\SU{N}{K+1}$.
$\Zb_{N}\times\Zb_{N}$ in \eqref{total_sym} is a subgroup of the center which acts trivially on the fermions.
When $K$ is even, this subgroup is generated by two elements
    \begin{align}
        c_{1}&:=(\Nsqrt,\,\Nsqrt,\,\dots,\,\Nsqrt,\,\Nsqrt,\,1),\\
        c_{2}&:=(\Nsqrt,\,1,\,\Nsqrt,\,1,\,\dots,\,\Nsqrt,\,1,\,\Nsqrt,\,\Nsqrt)\label{generator_c_two}.
    \end{align}
When $K$ is odd, on the other hand, it is generated by 
    \begin{align}
        c_{1}&=(\Nsqrt,\,\Nsqrt,\,\dots,\,\Nsqrt,\,\Nsqrt,\,1),\label{generator_c_one}\\
        c'_{2}&:=(\Nsqrt,\,1,\,\Nsqrt,\,1,\,\dots,\,\Nsqrt,\,1,\,\Nsqrt).
    \end{align}

\begin{figure}[thbp]
    \begin{minipage}[b]{0.49\textwidth}
    \centering
    \includegraphics[width=7cm]{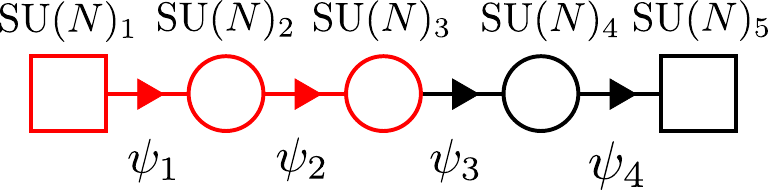}
  \end{minipage}
    \begin{minipage}[b]{0.49\textwidth}
    \centering
    \includegraphics[width=7cm]{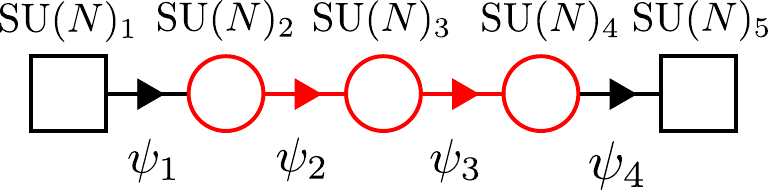}
  \end{minipage}
  \caption{Subquiver}\label{fig_subquiver}
\end{figure}

By considering subquivers, one finds that the symmetry is at most the group in \eqref{total_sym}. 
We focus on the fermions in the subquiver depicted by red arrowed lines in fig.~\ref{fig_subquiver}.
The symmetry of the original theory is a symmetry of the subquiver; in other words, the transformations that are not a symmetry of the subquiver are not a symmetry of the original theory.
In particular, the connected subquivers with three nodes correspond to massless QCD and its symmetry is well known.
For example, the symmetry of the subquiver in the left panel of fig.~\ref{fig_subquiver} is given by
    \begin{align}
        \frac{\SU{N}{1}\times\SU{N}{2}\times\SU{N}{3}\times\Uone}{\Zb_{N}\times\Zb_{N}}.
    \end{align}
In the same way, the subquiver theory in the right panel of fig.~\ref{fig_subquiver} has the symmetry
    \begin{align}
        \frac{\SU{N}{2}\times\SU{N}{3}\times\SU{N}{4}\times\Uone}{\Zb_{N}\times\Zb_{N}}.
    \end{align}
Since any subquiver with three nodes has such a symmetry, we conclude that the total symmetry is not larger than the group in \eqref{total_sym}.

To show that the total symmetry is just \eqref{total_sym}, we check that the $\Uone$ symmetry does not suffer from the Adler-Bell-Jackiw (ABJ) anomaly \cite{Adler:1969gk,Bell:1969ts}.
Actually, the fermion measure is not changed under the $\Uone$ transformation \eqref{uone_trans}:
    \begin{align}
    \mathcal{D}\bar{\psi}'\mathcal{D}\psi'
    &=\mathcal{D}\bar{\psi}\mathcal{D}\psi\exp\left(
        -\frac{\imaginary N\theta}{8\pi^{2}}\int\Tr f_{2}^{2}
        +\sum_{i=2}^{K-1}(-1)^{i}\frac{\imaginary N\theta}{8\pi^{2}}\int\left(\Tr f_{i}^{2}+\Tr f_{i+1}^{2}\right)
        +(-1)^{K}\frac{\imaginary N\theta}{8\pi^{2}}\int\Tr f_{K}^{2}
        \right)\notag\\
    &=\mathcal{D}\bar{\psi}\mathcal{D}\psi.
    \end{align}
Therefore, the liner quiver gauge theory has the symmetry \eqref{total_sym}.

By removing the gauge group from the total symmetry, we find that the global symmetry of the theory is
    \begin{align}\label{global_sym}
        G=\frac{\SU{N}{1}\times\SU{N}{K+1}\times\Uone}{\Zb_{N}\times\Zb_{N}}.
    \end{align}

\section{Vacuum structure of quiver theories with even number of edges}\label{sec_even}
In this section, we study the linear quiver gauge theory with even $K$ based on the anomaly matching condition.

\subsection{'t Hooft anomaly matching for even \texorpdfstring{$K$}{K}}\label{sub_even_zero}
We show that the 't~Hooft anomalies of the theory for even $K$ are exactly the same as those of massless QCD.
It is believed that chiral symmetry breaking occurs in massless QCD by the quark bilinear condensate.
In the present case, it seems to be natural that by the condensation of the gauge invariant operator\footnote{For $K\ge 4$, there are several operators with different ways to contract spinor indices in $\evenboson$ unlike massless QCD. However, it is difficult to distinguish them by some symmetries.}
    \begin{align}\label{even_boson}
        \evenboson\sim\psi_{1}\psi_{2}\cdots\psi_{K},
    \end{align}
the global symmetry \eqref{global_sym} is spontaneously broken as chiral symmetry.
The breaking pattern is
    \begin{align}\label{ssb_even}
        G=\frac{\SU{N}{1}\times\SU{N}{K+1}\times\Uone}{\Zb_{N}\times\Zb_{N}}\to H=\frac{\SU{N}{\diag}\times\Uone}{\Zb_{N}\times\Zb_{N}},
    \end{align}
where $\SU{N}{\diag}$ is the diagonal subgroup of $\SU{N}{1}\times\SU{N}{K+1}$.

We compute the anomaly polynomial to find the perturbative anomaly for our theory.
To that end, we introduce the $\SU{N}{1}$ gauge field $A_{1}$, the $\SU{N}{K+1}$ gauge field $A_{K+1}$ and the $\Uone$ gauge field $A$.
On the background fields, the covariant derivative on the fermions is given by
    \begin{align}
    D\psi_{1}&=\exdel\psi_{1}+A_{1}\psi_{1}-\psi_{1}a_{2}-A\psi_{1},\\
    D\psi_{i}&=\exdel\psi_{i}+a_{i}\psi_{i}-\psi_{i}a_{i+1}+(-1)^{i}A\psi_{i}\qquad(i=2,\,\dots,\,K-1),\\
    D\psi_{K}&=\exdel\psi_{K}+a_{K}\psi_{K}-\psi_{K}A_{K+1}+A\psi_{K+1}.
\end{align}
Then the $6$-dimensional anomaly polynomial of our theory is 
    \begin{align}
        I_{6}=\frac{\imaginary^{3}}{3!(2\pi^{3})}N\left(
            \Tr F_{1}^{3}-\Tr F_{K+1}^{3}+3\exdel A\Tr F_{1}^{2}-3\exdel A\Tr F_{K+1}^{2}
            \right).
    \end{align}
This is exactly the same as the anomaly polynomial of massless QCD.

Next we consider the anomaly-free subgroup 
    \begin{align}\label{even_free_subgroup}
        \frac{\SU{N}{\diag}}{\Zb_{N}}\times(\Zb_{N})_{\axial}\times\frac{\Uone}{\Zb_{N}}\subset G,
    \end{align}
to extract a discrete 't~Hooft anomaly.
The generator of $(\Zb_{N})_{\axial}$ acts on the fermions as
    \begin{align}\begin{aligned}\label{disc_chiral_trans_enen}
        \psi_{1}&\mapsto\psi_{1}'=\Nsqrt\psi_{1},\\
        \psi_{i}&\mapsto\psi_{i}'=\psi_{i}\qquad(i=2,\,\dots,\,K).
    \end{aligned}\end{align}
Then we introduce the $\SU{N}{\diag}$ gauge field $A_{\diag}$ and the $\Uone$ gauge field $A$.

We introduce two $\Zb_{N}$ 2-form gauge fields and formulate $\SU{N}{\diag}/\Zb_{N}$ and $\Uone/\Zb_{N}$ in \eqref{even_free_subgroup} following ref.~\cite{Kapustin:2014gua}.
For $\SU{N}{\diag}/\Zb_{N}$, we introduce a $2$-form $\Uone$ gauge field $B$ and a $1$-form $\Uone$ gauge field $C$ that satisfy the constraint
    \begin{align}\label{KS_constraint}
        NB=\exdel C.
    \end{align}
For $\Uone/\Zb_{N}$, we introduce another pair $(B',\,C')$ imposed the condition
    \begin{align}
        NB'=\exdel C'.
    \end{align}
Here the $\Uone$ gauge fields $B$, $B'$, $C$ and $C'$ are taken to be real though the gauge fields $a_{i}$ and $A_{\diag}$ for $0$-form symmetries are anti-hermitian.
In order to couple them to our theory, we promote the $\SU{N}{}$ gauge fields $a_{i}$ and $A_{\diag}$ to $\Unitary{N}{}$ gauge fields, locally written as
    \begin{align}
        \widetilde{A}_{\diag}&=A_{\diag}+\frac{\imaginary}{N}C1_{N}+\frac{\imaginary}{N}C'1_{N},\\
        \widetilde{a}_{2j+1}&=a_{2j+1}+\frac{\imaginary}{N}C1_{N}+\frac{\imaginary}{N}C'1_{N}\qquad(j=1,\,2,\,\dots,\,(K-2)/2),\\
        \widetilde{a}_{2j}&=a_{2j}+\frac{\imaginary}{N}C1_{N}\qquad(j=1,\,2,\,\dots,\,K/2).
    \end{align}
Then, we postulate the Abelian $1$-form gauge symmetries,
    \begin{align}\label{one_form_trans}
        B&\mapsto B+\exdel\lambda,&C&\mapsto C+N\lambda,
    \end{align}
and    
    \begin{align}\label{one_form_trans_prime}
        B'&\mapsto B'+\exdel\lambda',&C'&\mapsto C'+N\lambda',&A&\mapsto A+\imaginary\lambda',
    \end{align}
where $\lambda$ and $\lambda'$ are $\Uone$ gauge fields.
The covariant derivative that manifestly satisfies the invariance under the $1$-form gauge transformations is given by
    \begin{align}
    D\psi_{1}&=\exdel\psi_{1}+\widetilde{A}_{\diag}\psi_{1}-\psi_{1}\widetilde{a}_{2}-A\psi_{1},\\
    D\psi_{i}&=\exdel\psi_{i}+\widetilde{a}_{i}\psi_{i}-\psi_{i}\widetilde{a}_{i+1}+(-1)^{i}A\psi_{i}\qquad(i=2,\,\dots,\,K-1),\\
    D\psi_{K}&=\exdel\psi_{K}+\widetilde{a}_{K}\psi_{K}-\psi_{K}\widetilde{A}_{\diag}+A\psi_{K+1}.
\end{align}
We replace the field strengths by ones invariant under the transformations \eqref{one_form_trans} and \eqref{one_form_trans_prime}:
    \begin{align}
        F_{\diag}&\to\widetilde{F}_{\diag}-\imaginary B1_{N}-\imaginary B'1_{N},\\
        f_{2j+1}&\to\widetilde{f}_{2j+1}-\imaginary B1_{N}-\imaginary B'1_{N}\qquad(j=1,\,2,\,\dots,\,(K-2)/2),\\
        f_{2j}&\to\widetilde{f}_{2j+1}-\imaginary B1_{N}\qquad(j=1,\,2,\,\dots,\,K/2).
    \end{align}
The correctly normalized $\Uone/\Zb_{N}$ field strength is identified as
    \begin{align}
    N(\exdel A-\imaginary B').
    \end{align}
We can gauge the group
    \begin{align}
        \frac{\SU{N}{\diag}\times\SU{N}{2}\times\cdots\times\SU{N}{K}\times\Uone}{\Zb_{N}\times\Zb_{N}},
    \end{align}
by this procedure.

Finally we extract the discrete 't~Hooft anomaly.
To that end, we perform the $(\Zb_{N})_{\axial}$ transformation \eqref{disc_chiral_trans_enen}.
After the transformation, the fermion measure gets the phase
    \begin{align}
        &\exp\left(\frac{2\pi\imaginary}{N}\frac{\imaginary^{2}}{8\pi^{2}}\int\left(
            N\Tr(\widetilde{F}_{\diag}-\imaginary B1_{N}-\imaginary B'1_{N})^{2}+N\Tr(\widetilde{f}_{2}-\imaginary B1_{N})^{2}+N^{2}(\exdel A-\imaginary B')^{2}
            \right)\right)\notag\\
        &=\exp\left(-\frac{\imaginary N}{2\pi}\int (B\wedge B+B\wedge B')\right).
    \end{align}
This is the same as the discrete anomaly in massless QCD \cite{Tanizaki:2018wtg}.

Since the 't~Hooft anomalies of the linear quiver gauge theory for even $K$ coincide with the anomalies of massless QCD, it is natural that these theories behave in the same way at low energies.
Therefore, for the case of even $K$ we expect that the anomalies are matched by the spontaneous symmetry breaking \eqref{ssb_even}.
We now comment on the difference between massless QCD and the linear quiver theories for $K\ge 4$.
Since the theories for $K\ge 4$ are chiral unlike massless QCD, we cannot apply Vafa-Witten's theorem \cite{Vafa:1983tf} to them.
Thus the $\SU{N}{\diag}$ and $\Uone$ symmetries may be broken and it is possible for the global symmetry $G$ to be broken to a smaller group than the subgroup $H$ in \eqref{ssb_even}. 
For example, suppose that the global symmetry is broken as $G\to\Uone/\Zb_{N}$.
In this case, $2(N^{2}-1)$ Nambu-Goldstone bosons appear.
They are neutral under the remaining $\Uone/\Zb_{N}$ symmetry.
The anomalies are also matched by the Wess-Zumino-Witten term constructed by the procedure in Appendix C of ref.~\cite{Tachikawa:2016xvs}, since the $\Uone/\Zb_{N}$ group is an anomaly-free subgroup.
This possibility cannot be ruled out to the best of our knowledge.

\subsection{Thermal phase transition for even \texorpdfstring{$K$}{K}}\label{sub_even_thermal}
In this subsection, we consider the linear quiver gauge theory for even $K$ at finite temperature.
We introduce the imaginary chemical potential $\mu$ and see that at $\mu=\pi$ there is a $\Zbr$ symmetry which defines confinement following ref.~\cite{Yonekura:2019vyz}.
There is a mixed 't~Hooft anomaly between the $\Zbr$ symmetry and the global symmetry \eqref{global_sym}.
We argue the relation between the critical temperatures associated to the symmetries $\Zbr$ and \eqref{global_sym}, using the anomaly.

First, we define the thermal partition function with the imaginary chemical potential $\mu$.
To that end, we consider the subgroup
    \begin{align}
    \Ubaryon:=\frac{\Uone}{\Zb_{N}}\subset G.
    \end{align}
Since $\Ubaryon$ acts on the fermions as
    \begin{align}
        \psi_{i}\mapsto\psi'=\mathrm{e}^{\imaginary\theta(-1)^{i}/N}\psi_{i}\quad(i=1,\,\dots,\,K)
    \end{align}
with $\theta\sim\theta+2\pi$, gauge invariant operators have integer charges for the symmetry $\Ubaryon$.
The thermal partition function in the presence of $\mu$ is defined by
    \begin{align}\label{even_thermal_partition}
        Z(\beta,\,\mu):=\Tr\exp(-\beta H+\imaginary\mu Q_{B}),
    \end{align}
where $\beta$ is the inverse temperature, $H$ is the Hamiltonian and $Q_{B}$ is the charge operator for the $\Ubaryon$ symmetry.
We can obtain the same partition function by the Euclidean path integral on the product spacetime with the thermal circle $S^{1}_{\beta}$ of circumference $\beta$ and the spatial manifold $M_{3}$.
In the path integral formalism, the chemical potential is introduced as the holonomy of the background $\Ubaryon$ gauge field $A_{B}$ along the thermal circle $S^{1}_{\beta}$:
    \begin{align}
        \int_{S^{1}_{\beta}}A_{B}=-\imaginary\mu.
    \end{align}
We couple the theory on $S^{1}_{\beta}\times M_{3}$ to the background gauge field $A_{B}$ by the minimal coupling term $\exp(\imaginary\int_{S^{1}_{\beta}\times M_{3}}A_{B}\wedge\ast j_{B})$ with the current $j_{B}$ for the $\Ubaryon$ symmetry.

We define Wilson loops for fermions $\psi_{i}$ as
    \begin{align}\label{even_wilson_1}
        W_{1}&:=\mathrm{P}\exp\left(
        -\int_{S^{1}_{\beta}}\left(1_{N}\otimes a_{2}^{\ast}-\frac{1}{N}A_{B}1_{N}\otimes 1_{N}\right)
        \right)
        =\mathrm{e}^{-\imaginary\mu/N}\mathrm{P}\exp\left(
        -\int_{S^{1}_{\beta}}(1_{N}\otimes a_{2}^{\ast})
        \right),\\\label{even_wilson_i}
        W_{i}&:=\mathrm{P}\exp\left(
        -\int_{S^{1}_{\beta}}\left(a_{i}\otimes 1_{N}+1_{N}\otimes a_{i+1}^{\ast}+(-1)^{i}\frac{1}{N}A_{B}1_{N}\otimes 1_{N}\right)
        \right)\notag\\
        &=\mathrm{e}^{\imaginary\mu(-1)^{i}/N}\mathrm{P}\exp\left(
        -\int_{S^{1}_{\beta}}(a_{i}\otimes 1_{N}+1_{N}\otimes a_{i+1}^{\ast})
        \right)\qquad(i=2,\,\dots,\,K-1),\\\label{even_wilson_k}
        W_{K}&:=\mathrm{P}\exp\left(
        -\int_{S^{1}_{\beta}}\left(a_{K}\otimes 1_{1}+\frac{1}{N}A_{B}1_{N}\otimes 1_{N}\right)
        \right)
        =\mathrm{e}^{\imaginary\mu/N}\mathrm{P}\exp\left(
        -\int_{S^{1}_{\beta}}(a_{K}\otimes 1_{N})
        \right).
    \end{align}
We will see that these Wilson loops are order parameters for the confinement-deconfinement phase transition.

Let us consider the symmetries of the thermal partition function.
Note that the action of the subgroup $\Zb_{N}$ generated by the element $c_{2}$ in \eqref{generator_c_two} does not transform the fermions at all.
The thermal partition function is invariant under the $\Zb_{N}$ shift of the holonomies of gauge fields.
The chemical potential is changed to
    \begin{align}
        \int_{S^{1}_{\beta}}A_{B}=-\imaginary(\mu+2\pi)
    \end{align}
by the shift.
This means that the thermal partition function is invariant under the shift of the imaginary chemical potential as
    \begin{align}
        Z(\beta,\,\mu+2\pi)=Z(\beta,\,\mu).
    \end{align}
Moreover, there is a reflection symmetry $\mathcal{R}$ in the theory.
The transformation $\mathcal{R}$ is obtained by combining the time reversal and a reflection of one of the directions of the space $M_{3}$, and is given by
    \begin{align}
        \mathcal{R}:\,(x^{3},\,x^{4})\to(-x^{3},\,-x^{4}),
    \end{align}
where $x^{4}$ is a coordinate of $S^{1}_{\beta}$ and $x^{3}$ is one of the coordinates of $M_{3}$.
The reflection $\mathcal{R}$ transforms the chemical potential as
    \begin{align}
        \mathcal{R}:\,\mu\mapsto -\mu.
    \end{align}
Then the transformation $\mathcal{R}$ is a symmetry of the thermal partition function at $\mu=\pi$ since $\mu$ has periodicity of $2\pi$.
We denote the symmetry generated by $\mathcal{R}$ as $\Zbr$.

We see that the symmetry $\Zbr$ characterizes the confinement/deconfinement phase.
The Wilson loops \eqref{even_wilson_1}, \eqref{even_wilson_i}, and \eqref{even_wilson_k} transform as
    \begin{align}
        \mathcal{R}: W_{i}\mapsto W_{i}^{\dagger}.
    \end{align}
Note that the insertion of the Wilson loop $W_{i}$ corresponds to placing the fermion $\psi_{i}$ as the test particle along the thermal circle.
Since the $\Ubaryon$ charge of $\psi_{i}$ is $1/N$, in deconfinement phase the vacuum expectation value of the Wilson loop is given by
    \begin{align}
        \left\langle \Tr W_{i}\right\rangle\sim\exp\left(-\beta E+\frac{\imaginary\pi}{N}\right),
    \end{align}
where $E$ is the energy created by the test particle.
On the other hand, in confinement phase the vacuum expectation value is given by
    \begin{align}
        \left\langle\Tr W_{i}\right\rangle\sim\exp\left(-\beta E+\imaginary\pi m\right)\qquad(m\in\mathbb{Z}),
    \end{align}
since the test particle is combined with dynamical fermions to make a color singlet state.
Therefore we can use $\mathrm{Im}\left\langle\Tr W_{i}\right\rangle$ as a criterion of confinement/deconfinement.

Next we deduce the mixed 't~Hooft anomaly between $\Zbr$ and $\SU{N}{1}\times\SU{N}{K+1}$.
After the compactification on the thermal circle, we obtain the three dimensional effective theory.
In the effective theory, we regard the symmetry $\Zbr$ as the parity anomaly.
In order to compute the parity anomaly, we investigate the holonomies of the fermions along the thermal circle.
We can choose the specific configurations for the dynamical gauge fields $a_{i}$ in the argument since there is no anomaly involving them.
Here we employ the configurations which satisfy
    \begin{align}
        \mathrm{P}\exp\left(-\int_{S^{1}_{\beta}}a_{2n}\right)&=\begin{pmatrix}
        -\halfNsqinv&&&\\
        &\halfNsqinv&&\\
        &&\ddots&\\
        &&&\halfNsqinv
        \end{pmatrix}\qquad(n=1,\,2,\,\dots,\,K/2),\\
        \mathrm{P}\exp\left(-\int_{S^{1}_{\beta}}a_{2n+1}\right)&=1_{N}\qquad(n=1,\,2,\,\dots,\,K/2-1).
    \end{align}
Then the Wilson loops are given by
    \begin{align}
        W_{2n-1}&=1_{N}\otimes\begin{pmatrix}
        -1&&&\\
        &1&&\\
        &&\ddots&\\
        &&&1
        \end{pmatrix}\qquad(n=1,\,2,\,\dots,\,K/2),\\
        W_{2n}&=\begin{pmatrix}
        -1&&&\\
        &1&&\\
        &&\ddots&\\
        &&&1
        \end{pmatrix}\otimes 1_{N}\qquad(n=1,\,2,\,\dots,\,K/2).
    \end{align}
In addition to these holonomies, the fermions obtain the extra sign $(-1)$ from the spin structure since we consider the theory at finite temperature.
In particular, we need the boundary conditions of $\psi_{1}$ and $\psi_{K}$ on $S^{1}_{\beta}$ to compute the parity anomaly.
The only one kind of the fundamental fermions of $\SU{N}{1}$ among the fermions $\psi_{1}$ has the periodic boundary condition.
Similarly, a kind of the anti-fundamental fermions of $\SU{N}{K+1}$ among the fermions $\psi_{K}$ has the periodic boundary condition.
After we introduce the background gauge fields $A_{1}$ and $A_{K+1}$ for $\SU{N}{1}$ and $\SU{N}{K+1}$, by the parity transformation $\Zbr$ we obtain the parity anomaly \cite{Yonekura:2019vyz,Redlich:1983dv,Niemi:1983rq,Alvarez-Gaume:1984zst} given by
    \begin{align}\label{parity_anomaly}
        \exp\left\{\frac{1}{2}\left(\CS(A_{1})-\CS(A_{K+1})\right)\right\},
    \end{align}
where $\CS(A)$ is the Chern-Simons action:
    \begin{align}
        \CS(A):=-\frac{\imaginary}{4\pi}\int\Tr\left(
        A\exdel A+\frac{2}{3}A^{3}
        \right).
    \end{align}

Let us argue that we can obtain the same anomaly \eqref{parity_anomaly} from the theory at low temperature.
The effective theory after the spontaneous symmetry breaking \eqref{ssb_even} is the non-linear sigma model of Nambu-Goldstone bosons and the target space is given by the coset $G/H\simeq\SU{N}{}$.
In fact it is possible that we extract the anomaly \eqref{parity_anomaly} from the Wess-Zumino-Witten term similar to the case of massless QCD \cite{Yonekura:2019vyz} since the anomaly polynomial for the quiver theory is the same as massless QCD.
Thus the anomaly at low temperature is matched.

We discuss the phase structure of the theory using the above anomaly.
At high temperature the anomaly is matched by the symmetry breaking of $\Zbr$, while at low temperature it is matched by the symmetry breaking \eqref{ssb_even}.
The anomaly must be matched at middle temperature.
When both the symmetries $\Zbr$ and $\SU{N}{1}\times\SU{N}{K+1}$ are preserved, there must be some degrees of freedom to match the anomaly.
If we assume that there are no such exotic degrees of freedom, we conclude that either $\Zbr$ or $\SU{N}{1}\times\SU{N}{K+1}$ is broken.
Therefore the critical temperatures $\Tssb$ and $\Tconf$ associated to the symmetries $\Zbr$ and $\SU{N}{1}\times\SU{N}{K+1}$, respectively, satisfy the inequality
    \begin{align}
        \Tconf\le\Tssb.
    \end{align}

\section{Vacuum structure of quiver theories with odd number of edges}\label{sec_odd}
In this section, we consider the phase structure of the linear quiver gauge theory with odd $K$ by the anomaly matching condition.

\subsection{'t Hooft anomaly matching for odd \texorpdfstring{$K$}{K}}\label{sub_odd_zero}
We compute the 't~Hooft anomalies of the linear quiver gauge theory with odd $K$.
Note that in this case the symmetry breaking as massless QCD does not occur, since the gauge invariant operator $\psi_{1}\psi_{2}\cdots\psi_{K}$ in \eqref{even_boson} is now fermionic and cannot condense.
We show that the anomalies are matched by the massless composite fermions at low energy.
Such a possibility has been proposed in ref.~\cite{Georgi:1985hf}.
We also argue another possibility that the anomalies are matched by the Nambu-Goldstone bosons of spontaneous symmetry breaking as
    \begin{align}\label{ssb_odd}
        G=\frac{\SU{N}{1}\times\SU{N}{K+1}\times\Uone}{\Zb_{N}\times\Zb_{N}}\to
        \begin{cases}
        \frac{\SO{1}\times\SO{K+1}\times\Zb_{2}}{\Zb_{2}\times\Zb_{2}} & (N:\text{even})\\
        \SO{1}\times\SO{K+1}\times\Zb_{2} & (N:\text{odd}).
        \end{cases}
    \end{align}

First, we calculate the perturbative anomaly.
We introduce the $\SU{N}{1}$ gauge field $A_{1}$, the $\SU{N}{K+1}$ gauge field $A_{K+1}$ and the $\Uone$ gauge field $A$.
On these background fields, the covariant derivative on the fermions is given by
    \begin{align}
    D\psi_{1}&=\exdel\psi_{1}+A_{1}\psi_{1}-\psi_{1}a_{2}-A\psi_{1},\\
    D\psi_{i}&=\exdel\psi_{i}+a_{i}\psi_{i}-\psi_{i}a_{i+1}+(-1)^{i}A\psi_{i}\qquad(i=2,\,\dots,\,K-1),\\
    D\psi_{K}&=\exdel\psi_{K}+a_{K}\psi_{K}-\psi_{K}A_{K+1}-A\psi_{K+1}.
    \end{align}
Then, the $6$-dimensional anomaly polynomial is given by
    \begin{align}\label{eq_poly_odd}
        I_{6}=\frac{\imaginary^{3}}{3!(2\pi^{3})}N\left(
            \Tr F_{1}^{3}-\Tr F_{K+1}^{3}-N(\exdel A)^{3}-3\exdel A\Tr F_{1}^{2}-3\exdel A\Tr F_{K+1}^{2}
            \right).
    \end{align}

Next, we consider the anomaly-free subgroup
    \begin{align}\label{discsub_odd}
        \frac{\SU{N}{\diag}}{\Zb_{N}}\times(\Zb_{N})_{\axial}\subset G,
    \end{align}
where the generator of $(\Zb_{N})_{\axial}$ acts on the fermions as
    \begin{align}\begin{aligned}\label{disc_chiral_trans_odd}
        \psi_{1}&\mapsto\psi_{1}'=\Nsqinv\psi_{1},\\
        \psi_{i}&\mapsto\psi_{i}'=\psi_{i}\qquad(i=2,\,\dots,\,K).
    \end{aligned}\end{align}
In order to gauge the subgroup $\SU{N}{\diag}/\Zb_{N}$, we introduce the $1$-form $\SU{N}{\diag}$ gauge field $A_{\diag}$, the $2$-form $\Uone$ gauge field $B$ and the $1$-form $\Uone$ gauge field $C$.
These $\Uone$ gauge fields $B$ and $C$ are taken to be real and satisfy the constraint
    \begin{align}
        NB=\exdel C.
    \end{align}
Then we promote the $\SU{N}{}$ gauge fields $a_{i}$ and $A_{\diag}$ to $\Unitary{N}{}$ gauge fields, locally written as
    \begin{align}
        \widetilde{A}_{\diag}&=A_{\diag}+\frac{\imaginary}{N}C1_{N},\\
        \widetilde{a}_{i}&=a_{i}+\frac{\imaginary}{N}C1_{N}\qquad(i=2,\,\dots,\,K),
    \end{align}
and postulate the Abelian $1$-form gauge symmetry
    \begin{align}
        B&\mapsto B+\exdel\lambda,&C&\mapsto C+N\lambda.
    \end{align}
The covariant derivative that manifestly satisfies the invariance under the $1$-form gauge transformation is given by
    \begin{align}
    D\psi_{1}&=\exdel\psi_{1}+\widetilde{A}_{\diag}\psi_{1}-\psi_{1}\widetilde{a}_{2},\\
    D\psi_{i}&=\exdel\psi_{i}+\widetilde{a}_{i}\psi_{i}-\psi_{i}\widetilde{a}_{i+1}\qquad(i=2,\dots,\,K-1),\\
    D\psi_{K}&=\exdel\psi_{K}+\widetilde{a}_{K}\psi_{K}-\psi_{K}\widetilde{A}_{\diag}.
\end{align}
It is also necessary to replace the field strengths by ones invariant under the $1$-form gauge transformation:
    \begin{align}
        F_{\diag}&\to\widetilde{F}_{\diag}-\imaginary B1_{N},\\
        f_{i}&\to\widetilde{f}_{i}-\imaginary B1_{N}\qquad(i=2,\,\dots,\,K).
    \end{align}
We can gauge the group
    \begin{align}\label{discgaged_odd}
        \frac{\SU{N}{\diag}\times\SU{N}{2}\times\cdots\times\SU{N}{K}}{\Zb_{N}}
    \end{align}
by the above procedure.

Finally we perform the $(\Zb_{N})_{\axial}$ transformation \eqref{disc_chiral_trans_odd} on the background gauge fields to obtain the discrete 't~Hooft anomaly.
The fermion measure gets the phase
    \begin{align}\label{disc_anomaly_odd}
    &\exp\left(
    -\frac{2\pi \imaginary}{N}\frac{\imaginary^{2}}{8\pi^{2}}\int\left(
       N\Tr(\widetilde{F}_{\diag}-\imaginary B1_{N})^{2}+N\Tr(\widetilde{f}_{2}-\imaginary B1_{N})^{2} 
    \right)\right)\notag\\
    &=\exp\left(\frac{\imaginary N}{2\pi}\int B\wedge B\right),
    \end{align}
by the transformation \eqref{disc_chiral_trans_odd}.

We show that the anomalies deduced above are matched by the massless composite fermion
    \begin{align}
        \oddfermion\sim\psi_{1}\psi_{2}\cdots\psi_{K}
    \end{align}
at the infrared regime.
The operator $\oddfermion$ is in the bifundamental representation of the global symmetry $\SU{N}{1}\times\SU{N}{K+1}$ and has the charge $1$ under the $\Uone$ transformation.
Similar to the calculation at the ultraviolet regime, the covariant derivative of the fermion $\oddfermion$ on the background gauge field of $\SU{N}{1}\times\SU{N}{K+1}\times\Uone$ is given by
    \begin{align}
        D\oddfermion=\exdel\oddfermion+A_{1}\oddfermion-\oddfermion A_{K+1}-A\oddfermion.
    \end{align}
It is verified that the anomaly polynomial of the infrared theory is the same as that of the ultraviolet theory \eqref{eq_poly_odd}.
On the other hand, after gauging the group \eqref{discgaged_odd}, the covariant derivative is given by
    \begin{align}
        D\oddfermion=\exdel\oddfermion+\widetilde{A}_{\diag}\oddfermion-\oddfermion\widetilde{A}_{\diag}.
    \end{align}
Then we can also check that the phase which the fermion measure gets by the $(\Zb_{N})_{\axial}$ transformation $\oddfermion\mapsto\oddfermion'=\Nsqinv\oddfermion$ coincides with the phase \eqref{disc_anomaly_odd}.
Therefore we conclude that the anomalies are matched by the massless composite fermion $\oddfermion$.

As another possibility, we consider the case that the anomalies are matched by the Nambu-Goldstone bosons of spontaneous symmetry breaking of the global symmetry \eqref{global_sym}.
Note that the gauge invariant operator
    \begin{align}
        \oddboson\sim\psi_{1}\psi_{2}\cdots\psi_{K}\psi_{1}\psi_{2}\cdots\psi_{K}
    \end{align}
is in the bitensor representation of the group $\SU{N}{1}\times\SU{N}{K+1}$ and has the charge $2$ under the $\Uone$ transformation.
If the symmetric part of the operator $\oddboson$ condenses, the global symmetry $\SU{N}{1}\times\SU{N}{K+1}$ is broken to the subgroup $\SO{1}\times\SO{K+1}$.
Moreover, the $\Uone$ symmetry is broken to the subgroup $\Zb_{2}$ since the $\Uone$ charge of the operator $\oddboson$ is $2$.
As the result, the spontaneous symmetry breaking \eqref{ssb_odd} occurs.
Since the subgroup $\SO{1}\times\SO{K+1}$ is free from the perturbative anomaly, the perturbative anomaly of the global symmetry \eqref{global_sym} is matched by the Wess-Zumino-Witten term constructed by the procedure in Appendix C of ref.~\cite{Tachikawa:2016xvs}.
Note that the anomaly matching condition from the discrete anomaly \eqref{disc_anomaly_odd} requires the symmetry $\Uone$ to be broken  to the subgroup $\Zb_{2}$ or the trivial group.
The symmetry breaking \eqref{ssb_odd} also satisfies this condition.

\subsection{Thermal phase transition for odd \texorpdfstring{$K$}{K}}\label{sub_odd_thermal}
We consider the anomaly matching condition of the theory for odd $K$ at finite temperature.
In the case of the theory for \emph{even} $K$, we investigated the thermal partition function with imaginary chemical potential following ref.~\cite{Yonekura:2019vyz}.
In the present case, however, this method is not so effective since the symmetry and the expected behaviors at the infrared regime are different from those of massless QCD.
Instead, in what follows, we adopt a twisted thermal partition function introduced in ref.~\cite{Tanizaki:2017qhf}.

In order to define such a partition function, we introduce the $\SU{N}{1}$ gauge field $A_{1}$ and the $\SU{N}{K+1}$ gauge field $A_{K+1}$, and fix these holonomies along the thermal circle $S^{1}_{\beta}$:
    \begin{align}
    \mathrm{P}\exp\left(-\int_{S^{1}_{\beta}}A_{1}\right)&=\Omega\in\SU{N}{1},\\
    \mathrm{P}\exp\left(-\int_{S^{1}_{\beta}}A_{K+1}\right)&=\Omega\in\SU{N}{K+1},
    \end{align}
where
    \begin{align}
        \Omega:=\omega^{-(N-1)/2}\begin{pmatrix}
        1&&&&\\
        &\omega&&&\\
        &&\omega^{2}&&\\
        &&&\ddots&\\
        &&&&\omega^{N-1}
        \end{pmatrix},\qquad\omega:=\Nsqrt.
    \end{align}
Equivalently, this is the case that the twisted boundary conditions are imposed along $S^{1}_{\beta}$.
We then have the \emph{twisted} thermal partition function $Z(\beta,\,\Omega)$ of the theory coupled the above background gauge fields.

Let us verify the symmetry of the partition function $Z(\beta,\,\Omega)$.
Since the action of the subgroup $\Zb_{N}$ generated by the element $c_{1}$ in \eqref{generator_c_one} does not transform the fermions at all, the partition function is invariant under the $\Zb_{N}$ shift of the holonomies of the gauge fields as
    \begin{align}
        Z(\beta,\,\omega\Omega)=Z(\beta,\,\Omega).
    \end{align}
The holonomies of the dynamical gauge fields are also transformed as
    \begin{align}\label{themal_centertrans_odd}
        \mathrm{P}\exp\left(-\int_{S^{1}_{\beta}}a_{i}\right)\mapsto\omega\cdot\mathrm{P}\exp\left(-\int_{S^{1}_{\beta}}a_{i}\right)\qquad(i=2,\,3,\,\dots,\,K),
    \end{align}
by this $\Zb_{N}$ shift.
Besides the $\Zb_{N}$ symmetry, we perform the shift by the element $(S^{-1},\,S^{-1})\in\SU{N}{1}\times\SU{N}{K+1}$, where
    \begin{align}
        S=\begin{pmatrix}
        0&1&0&\cdots&0\\
        0&0&1&\dots&0\\
        \vdots&\vdots&\vdots&\ddots&\vdots\\
        0&0&0&\cdots&1\\
        1&0&0&\cdots&0
        \end{pmatrix}\in\SU{N}{}.
    \end{align}
This matrix $S$ satisfies the condition
    \begin{align}
        S\Omega S^{-1}=\omega\Omega.
    \end{align}
The transformation by the element $S$ is also symmetry of the partition function.
Note that the holonomies of the dynamical gauge fields are not changed by this shift while the holonomies of the background gauge fields are changed.

Now we combine the $\Zb_{N}$ shift with the transformation by the element $S$.
By this composition of the transformations, the holonomies of the background gauge fields are invariant, while those of the dynamical gauge fields are transformed as \eqref{themal_centertrans_odd}.
We denote the composition of the transformations as $\Zbc$.
Note that the background fields of $A_{1}$ and $A_{K+1}$ break the symmetry $\SU{N}{\diag}/\Zb_{N}$ in \eqref{discsub_odd} to the subgroup $\Uone^{N-1}/\Zb_{N}$.
In the following, we focus on the three symmetries, $\Zbc$, $\Uone^{N-1}/\Zb_{N}$ and $(\Zb_{N})_{\axial}$.

In order to gauge the symmetry groups $\Zbc$ and $\Uone^{N-1}/\Zb_{N}$, we introduce the $1$-form $\Zbc$ gauge field $B_{c_{1}}$, the $1$-form $\Uone^{N-1}$ gauge field $A_{T}$ and the $2$-form $\Zb_{N}$ gauge field $B_{T}$ on the spatial manifold $M_{3}$.
Using these background gauge fields, we can define the $\SU{N}{\diag}/\Zb_{N}$ gauge field on $S^{1}_{\beta}\times M_{3}$ as
    \begin{align}
        A&=A_{T}+\imaginary B_{c_{1}},\\
        B&=B_{T}+B_{c_{1}}\wedge\frac{1}{\beta}dx^{4},
    \end{align}
in addition to the background gauge fields $A_{1}$ and $A_{K+1}$, which provide the holonomy along $S^{1}_{\beta}$.
From the phase factor \eqref{disc_anomaly_odd} by the $(\Zb_{N})_{\axial}$ transformation, we obtain the anomaly at finite temperature on the background:
    \begin{align}\label{thermal_anomaly_odd}
        \exp\left(\frac{\imaginary N}{2\pi}\int_{M_{3}\times S^{1}_{\beta}}B^{2}\right)
        &=\exp\left(\frac{\imaginary N}{2\pi}\int_{M_{3}\times S^{1}_{\beta}}(B_{T}+B_{c_{1}}\wedge\frac{1}{\beta}dx^{4})^{2}\right)\notag\\
        &=\exp\left(\frac{\imaginary N}{\pi}\int_{M_{3}}B_{T}\wedge B_{c_{1}}\right).
    \end{align}

Finally, we argue the anomaly matching condition for \eqref{thermal_anomaly_odd}.
In subsection \ref{sub_odd_zero}, we propose two scenarios at the infrared regime for odd $K$.
At high temperature, the anomaly matching is satisfied by the spontaneous symmetry breaking of the $\Zbc$ symmetry in both cases.
If we assume that the infrared theory contains the massless composite fermions, the anomaly \eqref{thermal_anomaly_odd} is provided by \eqref{disc_anomaly_odd}.
On the other hand, if the global symmetry is spontaneously broken as \eqref{ssb_odd}, it is matched by the spontaneous symmetry breaking of the $(\Zb_{N})_{\axial}$ symmetry.

\subsection*{Acknowledgment}
We would like to thank Yui Hayashi, Naoto Kan, Masashi Kawahira, Yuta Nagoya, Ryosuke Sato, and Yuya Tanizaki for helpful discussions.
S.Y. would also like to thank the Yukawa Institute for Theoretical Physics at Kyoto University and Kavli Institute for the Physics and Mathematics of the Universe for hospitality during his stay. Discussions during the YITP workshop YITP-W-22-09 on ``Strings and Fields 2022'' were useful to complete this work.
This work was partially supported by Japan Society for the Promotion of Science (JSPS) Grant-in-Aid for Scientific Research Grant Numbers JP21J30003 (O.M.) and JP21K03574 (S.Y.).
\bibliographystyle{utphys}
\bibliography{ref}
\end{document}